\documentclass[english]{article}
\usepackage[T1]{fontenc}
\usepackage[latin9]{inputenc}
\usepackage{geometry}
\usepackage{color}
\usepackage{babel}
\usepackage{float}
\usepackage{amsmath}
\usepackage{graphicx}
\usepackage[unicode=true,
 bookmarks=false,
 breaklinks=false,pdfborder={0 0 1},backref=section,colorlinks=true]
 {hyperref}
\hypersetup{
 citecolor=blue}
\usepackage{breakurl}

\makeatletter

\floatstyle{ruled}
\newfloat{algorithm}{tbp}{loa}
\providecommand{\algorithmname}{Algorithm}
\floatname{algorithm}{\protect\algorithmname}

\usepackage{babel}

\renewcommand{\@biblabel}[1]{}
\renewenvironment{thebibliography}[1]
     {\section*{\refname}%
      \@mkboth{\MakeUppercase\refname}{\MakeUppercase\refname}%
      \list{\@biblabel{\@arabic\c@enumiv}}%
           {\settowidth\labelwidth{\@biblabel{#1}}%
            \leftmargin\labelwidth
            \advance\leftmargin\labelsep
            \itemindent-\labelsep
            \@openbib@code
            \usecounter{enumiv}%
            \let\p@enumiv\@empty
            \renewcommand{\theenumiv}{\@arabic\c@enumiv}}%
      \sloppy
      \clubpenalty4000
      \@clubpenalty \clubpenalty
      \widowpenalty4000%
      \sfcode`\.\@m}
     {\def\@noitemerr
       {\@latex@warning{Empty `thebibliography' environment}}%
      \endlist}

\usepackage{cite}

\usepackage{indentfirst}
\usepackage{algorithm}
\usepackage{algpseudocode}

\usepackage{babel}

\makeatother

\begin{document}

\title{The deterministic information bottleneck}

\author{DJ Strouse\thanks{Department of Physics, Princeton University}~
\& David J. Schwab\thanks{Department of Physics, Northwestern University}}
\maketitle
\begin{abstract}
Lossy compression and clustering fundamentally involve a decision
about what features are relevant and which are not. The information
bottleneck method (IB) by Tishby, Pereira, and Bialek formalized this
notion as an information-theoretic optimization problem and proposed
an optimal tradeoff between throwing away as many bits as possible,
and selectively keeping those that are most important. In the IB,
compression is measure my mutual information. Here, we introduce an
alternative formulation that replaces mutual information with entropy,
which we call the deterministic information bottleneck (DIB), that
we argue better captures this notion of compression. As suggested
by its name, the solution to the DIB problem turns out to be a deterministic
encoder, or hard clustering, as opposed to the stochastic encoder,
or soft clustering, that is optimal under the IB. We compare the IB
and DIB on synthetic data, showing that the IB and DIB perform similarly
in terms of the IB cost function, but that the DIB significantly outperforms
the IB in terms of the DIB cost function. We also empirically find
that the DIB offers a considerable gain in computational efficiency
over the IB, over a range of convergence parameters. Our derivation
of the DIB also suggests a method for continuously interpolating between
the soft clustering of the IB and the hard clustering of the DIB.
\end{abstract}

\section{Introduction}

\begin{sloppypar}Compression is a ubiquitous task for humans and
machines alike \cite{Cover,MacKay}. For example, machines must turn
the large pixel grids of color that form pictures into small files
capable of being shared quickly on the web \cite{JPEG}, humans must
compress the vast stream of ongoing sensory information they receive
into small changes in the brain that form memories \cite{Kandel},
and data scientists must turn large amounts of high-dimensional and
messy data into more manageable and interpretable clusters \cite{MacKay}.\end{sloppypar}

\begin{sloppypar}Lossy compression involves an implicit decision
about what is relevant and what is not \cite{Cover,MacKay}. In the
example of image compression, the algorithms we use deem some features
essential to representing the subject matter well, and others are
thrown away. In the example of human memory, our brains deem some
details important enough to warrant attention, and others are forgotten.
And in the example of data clustering, information about some features
is preserved in the mapping from data point to cluster ID, while information
about others is discarded.\end{sloppypar}

In many cases, the criterion for ``relevance'' can be described
as information about some other variable(s) of interest. Let's call
$X$ the signal we are compressing, $T$ the compressed version, $Y$
the other variable of interest, and $I\!\left(T;Y\right)$ the ``information''
that $T$ has about $Y$ (we will formally define this later). For
human memory, $X$ is past sensory input, $T$ the brain's internal
representation (e.g.~the activity of a neural population, or the
strengths of a set of synapses), and $Y$ the features of the future
environment that the brain is interested in predicting, such as extrapolating
the position of a moving object. Thus, $I\!\left(T;Y\right)$ represents
the predictive power of the memories formed \cite{Palmer}. For data
clustering, $X$ is the original data, $T$ is the cluster ID, and
$Y$ is the target for prediction, for example purchasing or ad-clicking
behavior in a user segmentation problem. In summary, a good compression
algorithm can be described as a tradeoff between the compression of
a signal and the selective maintenance of the ``relevant'' bits
that help predict another signal.

This problem was formalized as the ``information bottleneck'' (IB)
by Tishby, Pereira, and Bialek \cite{IB}. Their formulation involved
an information-theoretic optimazation problem, and resulted in an
iterative soft clustering algorithm guaranteed to converge to a local
(though not necessarily global) optimum. In their cost functional,
compression was measured by the mutual information $I\!\left(X;T\right)$.
This compression metric has its origins in rate-distortion theory
and channel coding, where $I\!\left(X;T\right)$ represents the maximal
information transfer rate, or capacity, of the communication channel
between $X$ and $T$ \cite{Cover}. While this approach has its applications,
often one is more interested in directly restricting the amount of
resources required to represent $T$, represented by the entropy $H\!\left(T\right)$.
This latter notion comes from the source coding literature and implies
a restriction on the \emph{representational cost} of $T$ \cite{Cover}.
In the case of human memory, for example, $H\!\left(T\right)$ would
roughly correspond to the number of neurons or synapses required to
represent or store a sensory signal $X$. In the case of data clustering,
$H\!\left(T\right)$ is related to the number of clusters.

In the following paper, we introduce an alternative formulation of
the IB, called the deterministic information bottleneck (DIB), replacing
the compression measure $I\!\left(X;T\right)$ with $H\!\left(T\right)$,
thus emphasizing contraints on representation, rather than communication.
Using a clever generalization of both cost functionals, we derive
an iterative solution to the DIB, which turns out to provide a hard
clustering, or deterministic mapping from $X$ to $T$, as opposed
to the soft clustering, or probabilitic mapping, that IB provides.
Finally, we compare the IB and DIB solutions on synthetic data to
help illustrate their differences.

\section{The original information bottleneck (IB)}

Given the joint distribution $p\!\left(x,y\right)$, the encoding
distribution $q\!\left(t|x\right)$ is obtained through the following
``information bottleneck'' (IB) optimization problem: 
\begin{align}
\min_{q\left(t|x\right)}L\!\left[q\!\left(t|x\right)\right] & =I\!\left(X;T\right)-\beta I\!\left(Y;T\right),\label{eq:IB_cost}
\end{align}

subject to the Markov constraint $T\leftrightarrow X\leftrightarrow Y$.
Here $I\!\left(X;T\right)$ denotes the mutual information between
$X$ and $T$, that is $I\!\left(X;T\right)\equiv H\!\left(T\right)-H\!\left(T|X\right)=\sum_{x,t}p\!\left(x,t\right)\log\!\left(\frac{p\left(x,t\right)}{p\left(x\right)p\left(t\right)}\right)=D_{KL}\!\left[p\!\left(x,t\right)\mid p\!\left(x\right)p\!\left(t\right)\right]$,\footnote{Implicit in the summation here, we have assumed that $X$, $Y$, and
$T$ are discrete. We will be keeping this assumption throughout for
convenience of notation, but note that the IB generalizes naturally
to $X$, $Y$, and $T$ continuous by simply replacing the sums with
integrals (see, for example, \cite{GIB}).} where $D_{KL}$ denotes the Kullback-Leibler divergence.\footnote{For those unfamiliar with it, mutual information is a very general
measure of how related two variables are. Classic correlation measures
typically assume a certain form of the relationship between two variables,
say linear, whereas mutual information is agnostic as to the details
of the relationship. One intuitive picture comes from the entropy
decomposition: $I\!\left(X;Y\right)\equiv H\!\left(X\right)-H\!\left(X|Y\right)$.
Since entropy measures uncertainty, mutual information measures the
\emph{reduction in uncertainty} in one variable when observing the
other. Moreover, it is symmetric ($I\!\left(X;Y\right)=I\!\left(Y;X\right)$),
so the information is \emph{mutual}. Another intuitive picture comes
from the $D_{KL}$ form: $I\!\left(X;Y\right)\equiv D_{KL}\!\left[p\!\left(x,y\right)\mid p\!\left(x\right)p\!\left(y\right)\right]$.
Since $D_{KL}$ measures the distance between two probability distributions,
the mutual information quantifies how far the relationship between
$x$ and $y$ is from a probabilistically independent one, that is
how far the joint $p\!\left(x,y\right)$ is from the factorized $p\!\left(x\right)p\!\left(y\right)$.
A very nice summary of mutual information as a dependence measure
is included in \cite{Kinney}.} The first term in the cost function is meant to encourage compression,
while the second relevance. $\beta$ is a non-negative free parameter
representing the relative importance of compression and relevance,
and our solution will be a function of it. The Markov constraint simply
enforces the probabilistic graphical structure of the task; the compressed
representation $T$ is a (possibly stochastic) function of $X$ and
can only get information about $Y$ through $X$. Note that we are
using $p$ to denote distributions that are given and fixed, and $q$
to denote distributions that we are free to change and that are being
updated throughout the optimization process.

Through a standard application of variational calculus (see Section~\ref{sec:Derivation}
for a detailed derivation of the solution to a more general problem
introduced below), one finds the formal solution:\footnote{For the reader familiar with rate-distortion theory, eqn~\ref{eq:IB_encoding}
can be viewed as the solution to a rate-distortion problem with distortion
measure given by the KL-divergence term in the exponent.}
\begin{align}
q\!\left(t|x\right) & =\frac{q\!\left(t\right)}{Z\!\left(x,\beta\right)}\exp\!\left[-\beta D_{\text{KL}}\!\left[p\!\left(y|x\right)\mid q\!\left(y|t\right)\right]\right]\label{eq:IB_encoding}\\
q\!\left(y|t\right) & =\frac{1}{q\!\left(t\right)}\sum_{x}q\!\left(t|x\right)p\!\left(x,y\right),\label{eq:qyt}
\end{align}

where $Z\!\left(x,\beta\right)\equiv\exp\!\left[-\frac{\lambda\left(x\right)}{p\left(x\right)}-\beta\sum_{y}p\!\left(y\mid x\right)\log\frac{p\left(y\mid x\right)}{p\left(y\right)}\right]$
is a normalization factor, and $\lambda\!\left(x\right)$ is a Lagrange
multiplier that enters enforcing normalization of $q\!\left(t\mid x\right)$.\footnote{More explicitly, our cost function $L$ also implicitly includes a
term $\sum_{x}\lambda\!\left(x\right)\left[1-\sum_{t}q\!\left(t|x\right)\right]$
and this is where $\lambda\!\left(x\right)$ comes in to the equation.
See Section~\ref{sec:Derivation} for details.} To get an intuition for this solution, it is useful to take a clustering
perspective - since we are compressing $X$ into $T$, many $X$ will
be mapped to the same $T$ and so we can think of the IB as ``clustering''
$x$s into their cluster labels $t$. The solution $q\!\left(t|x\right)$
is then likely to map $x$ to $t$ when $D_{\text{KL}}\!\left[p\!\left(y|x\right)\mid q\!\left(y|t\right)\right]$
is small, or in other words, when the distributions $p\!\left(y|x\right)$
and $q\!\left(y|t\right)$ are similar. These distributions are similar
to the extent that $x$ and $t$ provide similar information about
$y$. In summary, inputs $x$ get mapped to clusters $t$ that maintain
information about $y$, as was desired.

This solution is ``formal'' because the first equation depends on
the second and vice versa. However, \cite{IB} showed that an iterative
approach can be built on the the above equations which provably converges
to a local optimum of the IB cost function (eqn.~\ref{eq:IB_cost}).

Starting with some initial distributions $q^{\left(0\right)}\!\left(t|x\right)$,
$q^{\left(0\right)}\!\left(t\right)$, and $q^{\left(0\right)}\!\left(y|t\right)$,
the $n^{\text{th}}$ update is given by:\footnote{\label{fn:decreasingclusters-1}Note that, if at step $m$ no $x$s
are assigned to a particular $t=t^{*}$ (i.e. $q\!\left(t\mid x\right)=0\,\,\forall x$),
then $q^{\left(m\right)}\!\left(t^{*}\right)=q^{\left(m+1\right)}\!\left(t^{*}\right)=0$.
That is, no $x$s will ever again be assigned to $t^{*}$ (due to
the $q^{\left(n-1\right)}\!\left(t\right)$ factor in $q^{\left(n-1\right)}\!\left(t\mid x\right)$).
In other words, the number of $t$s ``in use'' can only decrease
during the iterative algorithm (or remain constant). Thus, it seems
plausible that the solution will not depend on the cardinality of
$T$, provided it is chosen to be large enough.}

\begin{align}
q^{\left(n\right)}\!\left(t|x\right) & =\frac{q^{\left(n-1\right)}\!\left(t\right)}{Z^{\left(n\right)}\!\left(x,\beta\right)}\exp\!\left[-\beta D_{\text{KL}}\!\left[p\!\left(y|x\right)\mid q^{\left(n-1\right)}\!\left(y|t\right)\right]\right]\\
q^{\left(n\right)}\!\left(t\right) & =\sum_{x}q^{\left(n\right)}\!\left(t|x\right)p\!\left(x\right)\\
q^{\left(n\right)}\!\left(y|t\right) & =\frac{1}{q^{\left(n\right)}\!\left(t\right)}\sum_{x}q^{\left(n\right)}\!\left(t|x\right)p\!\left(x,y\right).
\end{align}

Note that the first equation is the only ``meaty'' one; the other
two are just there to enforce consistency with the laws of probability
(e.g.~that marginals are related to joints as they should be). In
principle, with no proof of convergence to a global optimum, it might
be possible for the solution obtained to vary with the initialization,
but in practice, the cost function is ``smooth enough'' that this
does not seem to happen. This algorithm is summarized in algorithm~\ref{alg:IB}.
Note that while the general solution is iterative, there is at least
one known case in which an analytic solution is possible, namely when
$X$ and $Y$ are jointly Gaussian \cite{GIB}.

\begin{algorithm}[H]
\begin{algorithmic}[1]
\State{Given $p\!\left(x,y\right)$, $\beta \geq 0$}
\State{Initialize $q^{\left(0\right)}\!\left(t\mid x\right)$}
\State{$q^{\left(0\right)}\!\left(t\right)=\sum_{x}p\!\left(x\right)q^{\left(0\right)}\!\left(t\mid x\right)$}
\State{$q^{\left(0\right)}\!\left(y\mid t\right)=\frac{1}{q^{\left(0\right)}\!\left(t\right)}\sum_{x}p\!\left(x,y\right)q^{\left(0\right)}\!\left(t\mid x\right)$}
\State{$n=0$}
\While{not converged}
\State{$n=n+1$}
\State{$q^{\left(n\right)}\!\left(t\mid x\right)=\frac{q^{\left(n-1\right)}\!\left(t\right)}{Z\!\left(x,\beta\right)}\exp\!\left[-\beta D_{\text{KL}}\!\left[p\!\left(y\mid x\right)\mid q^{\left(n-1\right)}\!\left(y\mid t\right)\right]\right]$} \State{$q^{\left(n\right)}\!\left(t\right)=\sum_{x}p\!\left(x\right)q^{\left(n\right)}\!\left(t\mid x\right)$}
\State{$q^{\left(n\right)}\!\left(y\mid t\right)=\frac{1}{q^{\left(n\right)}\!\left(t\right)}\sum_{x}q^{\left(n\right)}\!\left(t\mid x\right)p\!\left(x,y\right)$}
\EndWhile
\end{algorithmic}

\caption{\textbf{- The information bottleneck (IB) method}.\label{alg:IB}}
\end{algorithm}

\begin{figure}[H]
\begin{centering}
\includegraphics[scale=0.3]{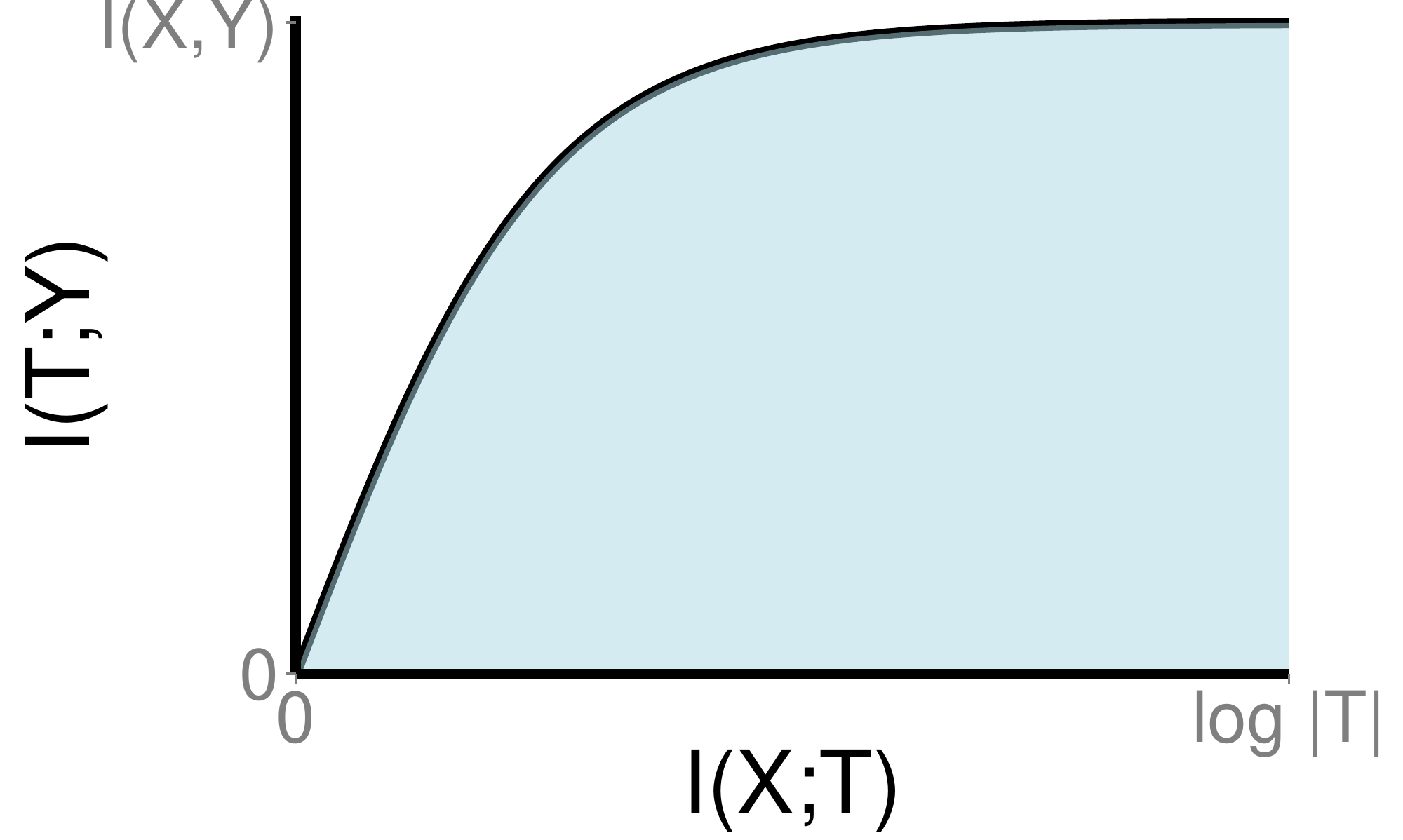} 
\par\end{centering}
\caption{\textbf{An illustrative IB curve.\label{fig:IB-curve}} $I\!\left(T;Y\right)$
is the relevance term from eqn~\ref{eq:IB_cost}; $I\!\left(X;T\right)$
is the compression term. $I\!\left(X;Y\right)$ is an upper bound
on $I\!\left(T;Y\right)$ since $T$ only gets its information about
$Y$ via $X$. $\log\!\left(\left|T\right|\right)$, where $\left|T\right|$
is the cardinality of the compression variable, is a bound on $I\!\left(X;T\right)$
since $I\!\left(X;T\right)=H\!\left(T\right)-H\!\left(T\mid X\right)\leq H\!\left(T\right)\leq\log\!\left(\left|T\right|\right)$.}
\end{figure}

In summary, given the joint distribution $p\!\left(x,y\right)$, the
IB method extracts a compressive encoder $q\!\left(t\mid x\right)$
that selectively maintains the bits from $X$ that are informative
about $Y$. As the encoder is a function of the free parameter $\beta$,
we can visualize the entire family of solutions on a curve (figure~\ref{fig:IB-curve}),
showing the tradeoff between compression (on the $x$-axis) and relevance
(on the $y$-axis), with $\beta$ as an implicitly varying parameter.
For small $\beta$, compression is more important than prediction
and we find ourselves at the bottom left of the curve in the high
compression, low prediction regime. As $\beta$ increases, prediction
becomes more important relative to compression, and we see that both
$I\!\left(X;T\right)$ and $I\!\left(T;Y\right)$ increase. At some
point, $I\!\left(T;Y\right)$ saturates, because there is no more
information about $Y$ that can be extracted from $X$ (either because
$I\!\left(T;Y\right)$ has reached $I\!\left(X;Y\right)$ or because
$T$ has too small cardinality). In this regime, the encoder will
approach the trivially deterministic solution of mapping each $x$
to its own cluster. At any point on the curve, the slope is equal
to $\beta^{-1}$, which we can read off directly from the cost functional.
Note that the region below the curve is shaded because this area is
feasible; for suboptimal $q\!\left(t\mid x\right)$, solutions will
lie in this region. Optimal solutions will of course lie on the curve,
and no solutions will lie above the curve.

\begin{sloppypar}Additional work on the IB has highlighted its relationship
with maximum likelihood on a multinomial mixture model \cite{ML-IB}
and canonical correlation analysis \cite{CCA-IB} (and therefore linear
Gaussian models \cite{CCA-LG} and slow feature analysis \cite{SFA-LG}).
Applications have included speech recognition \cite{Speech1,Speech2,Speech3},
topic modeling \cite{Topic1,Topic2,Topic3,Topic4}, and neural coding
\cite{Neural,Palmer}. Most recently, the IB has even been proposed
as a method for benchmarking the performance of deep neural networks
\cite{IB-DNN}.\end{sloppypar}

\section{The deterministic information bottleneck\label{sec:DIB}}

Our motivation for introducing an alternative formulation of the information
bottleneck is rooted in the ``compression term'' of the IB cost
function; there, the minimization of the mutual information $I\!\left(X;T\right)$
represents compression. As discussed above, this measure of compression
comes from the channel coding literature and implies a restriction
on the \emph{communication cost} between $X$ and $T.$ Here, we are
interested in the source coding notion of compression, which implies
a restriction on the \emph{representational cost} of $T$. For example,
in neuroscience, there is a long history of work on ``redundancy
reduction'' in the brain in the form of minimizing $H\!\left(T\right)$
\cite{Barlow1,Barlow2,Barlow3}.

Let us call the original IB cost function $L_{\text{IB}}$, that is
$L_{\text{IB}}\equiv I\!\left(X;T\right)-\beta I\!\left(T;Y\right)$.
We now introduce the deterministic information bottleneck (DIB) cost
function:
\begin{align}
L_{\text{DIB}}\!\left[q\!\left(t|x\right)\right] & \equiv H\!\left(T\right)-\beta I\!\left(T;Y\right),
\end{align}

which is to be minimized over $q\!\left(t\mid x\right)$ and subject
to the same Markov constraint as the original formulation (eqn~\ref{eq:IB_cost}).
The motivation for the ``deterministic'' in its name will become
clear in a moment.

To see the distinction between the two cost functions, note that:
\begin{align}
L_{\text{IB}}-L_{\text{DIB}} & =I\!\left(X;T\right)-H\!\left(T\right)\\
 & =-H\!\left(T\mid X\right),
\end{align}

where we have used the decomposition of the mutual information $I\!\left(X;T\right)=H\!\left(T\right)-H\!\left(T\mid X\right)$.
$H\!\left(T\mid X\right)$ is sometimes called the ``noise entropy''
and measures the stochasticity in the mapping from $X$ to $T$. Since
we are minimizing these cost functions, this means that the IB cost
function \emph{encourages} stochasticity in the encoding distribution
$q\!\left(t\mid x\right)$ relative to the DIB cost function. In fact,
we will see that by removing this encouragement of stochasticity,
the DIB problem actually produces a deterministic encoding distribution,
i.e. an encoding \emph{function}, hence the ``deterministic'' in
its name.

Naively taking the same variational calculus approach as for the IB
problem, one cannot solve the DIB problem.\footnote{When you take the variational derivative of $L_{\text{DIB}}+\text{Lagrange multiplier term}$
with respect to $q\!\left(t\mid x\right)$ and set it to zero, you
get no explicit $q\!\left(t\mid x\right)$ term, and it is therefore
not obvious how to solve these equations. We cannot rule that that
approach is possible, of course; we have just here taken a different
route.} To make this problem tractable, we are going to define a family of
cost functions of which the IB and DIB cost functions are limiting
cases. That family, indexed by $\alpha$, is defined as:\footnote{Note that for $\alpha<1$, we cannot allow $T$ to be continuous since
$H\!\left(T\right)$ can become infinitely negative, and the optimal
solution in that case will trivially be a delta function over a single
value of $T$ for all $X$, across all values of $\beta$. This is
in constrast to the IB, which can handle continuous $T$. In any case,
we continue to assume discrete $X$, $Y$, and $T$ for convenience.} 
\begin{align}
L_{\alpha} & \equiv H\!\left(T\right)-\alpha H\!\left(T\mid X\right)-\beta I\!\left(T;Y\right).
\end{align}

Clearly, $L_{\text{IB}}=L_{1}$. However, instead of looking at $L_{\text{DIB}}$
as the $\alpha=0$ case, we'll define the DIB solution $q_{\text{DIB}}\!\left(t\mid x\right)$
as the $\alpha\rightarrow0$ limit of the solution to the generalized
problem $q_{\alpha}\!\left(t\mid x\right)$:\footnote{Note a subtlety here that we cannot claim that the $q_{\text{DIB}}$
is the solution to $L_{\text{DIB}}$, for although $L_{\text{DIB}}=\lim_{\alpha\rightarrow0}L_{\alpha}$
and $q_{\text{DIB}}=\lim_{\alpha\rightarrow0}q_{\alpha}$, the solution
of the limit need not be equal to the limit of the solution. It would,
however, be surprising if that were not the case.} 
\begin{align}
q_{\text{DIB}}\!\left(t\mid x\right) & \equiv\lim_{\alpha\rightarrow0}q_{\alpha}\!\left(t\mid x\right).
\end{align}

Taking the variational calculus approach to minimizing $L_{\alpha}$
(under the Markov constraint), we get the following solution for the
encoding distribution (see Section~\ref{sec:Derivation} for the
derivation and explicit form of the normalization factor $Z\!\left(x,a,\beta\right)$):
\begin{align}
q_{\alpha}\!\left(t|x\right) & =\frac{1}{Z\!\left(x,\alpha,\beta\right)}\exp\!\left[\frac{1}{\alpha}\left(\log q_{\alpha}\!\left(t\right)-\beta D_{\text{KL}}\!\left[p\!\left(y|x\right)\mid q_{\alpha}\!\left(y|t\right)\right]\right)\right]\label{eq:qtx}\\
q_{\alpha}\!\left(y|t\right) & =\frac{1}{q_{\alpha}\!\left(t\right)}\sum_{x}p\!\left(y|x\right)q_{\alpha}\!\left(t|x\right)p\!\left(x\right).\label{eq:qyt2}
\end{align}

Note that the last equation is just eqn~\ref{eq:qyt}, since this
just follows from the Markov constraint. With $\alpha=1$, we can
see that the first equation just becomes the IB solution from eqn~\ref{eq:IB_encoding},
as should be the case.

Before we take the $\alpha\rightarrow0$ limit, note that we can now
write a generalized IB iterative algorithm (indexed by $\alpha$)
which includes the original as a special case ($\alpha=1$):
\begin{align}
q_{\alpha}^{\left(n\right)}\!\left(t|x\right) & =\frac{1}{Z\!\left(x,\alpha,\beta\right)}\exp\!\left[\frac{1}{\alpha}\left(\log q_{\alpha}^{\left(n-1\right)}\!\left(t\right)-\beta D_{\text{KL}}\!\left[p\!\left(y|x\right)\mid q_{\alpha}^{\left(n-1\right)}\!\left(y|t\right)\right]\right)\right]\\
q_{\alpha}^{\left(n\right)}\!\left(t\right) & =\sum_{x}p\!\left(x\right)q_{\alpha}^{\left(n\right)}\!\left(t|x\right)\\
q_{\alpha}^{\left(n\right)}\!\left(y|t\right) & =\frac{1}{q_{\alpha}^{\left(n\right)}\!\left(t\right)}\sum_{x}q_{\alpha}^{\left(n\right)}\!\left(t|x\right)p\!\left(x,y\right).
\end{align}

This generalized algorithm can be used in its own right, however we
will not discuss that option further here.

For now, we take the limit $\alpha\rightarrow0$ and see that something
interesting happens with $q_{\alpha}\!\left(t\mid x\right)$ - the
argument of the exponential begins to blow up. For a fixed $x$, this
means that $q\!\left(t\mid x\right)$ will collapse into a delta function
at the value of $t$ which maximizes $\log q\!\left(t\right)-\beta D_{\text{KL}}\!\left[p\!\left(y\mid x\right)\mid q\!\left(y\mid t\right)\right]$.
That is:
\begin{align}
\lim_{\alpha\rightarrow0}q_{\alpha}\!\left(t|x\right) & =f:X\rightarrow T,
\end{align}

where: 
\begin{equation}
f\!\left(x\right)=t^{*}=\text{\ensuremath{\underset{t}{\operatorname{argmax}}}}\left(\log q\!\left(t\right)-\beta D_{\text{KL}}\!\left[p\!\left(y|x\right)\mid q\!\left(y|t\right)\right]\right).\label{eq:DIB_encoder}
\end{equation}

So, as anticipated, the solution to the DIB problem is a deterministic
encoding distribution. The $\log q\!\left(t\right)$ above encourages
that we use as few values of $t$ as possible, via a ``rich-get-richer''
scheme that assigns each $x$ preferentially to a $t$ already with
many $x$s assigned to it. The $\text{KL}$ divergence term, as in
the original IB problem, just makes sure we pick $t$s which retain
as much information from $x$ about $y$ as possible. The parameter
$\beta$, as in the original problem, controls the tradeoff between
how much we value compression and prediction.

Also like in the original problem, the solution above is only a formal
solution, since eqn~\ref{eq:qtx} depends on eqn~\ref{eq:qyt2}
and vice versa. So we will again need to take an iterative approach;
in analogy to the IB case, we repeat the following updates to convergence
(from some initialization):\footnote{\label{fn:decreasingclusters-2}As with the IB, the DIB has the property
that once a cluster goes unused, it will not be brought back into
use in future steps. That is, if $q^{\left(m\right)}\!\left(t\right)=0$,
then $\log q^{\left(m\right)}\!\left(t\right)=-\infty$ and $q^{\left(m+1\right)}\!\left(t\mid x\right)=0\,\,\forall x$.
So once again, one should conservatively choose the cardinality of
$T$ to be ``large enough''; for both the IB and DIB, we chose to
set it equal to the cardinality of $X$.}

\begin{align}
f^{\left(n\right)}\!\left(x\right) & =\underset{t}{\text{argmax}}\!\left(\log q^{\left(n-1\right)}\!\left(t\right)-\beta D_{\text{KL}}\!\left[p\!\left(y|x\right)\mid q^{\left(n-1\right)}\!\left(y|t\right)\right]\right)\\
q^{\left(n\right)}\!\left(t|x\right) & =\delta\!\left(t-f^{\left(n\right)}\!\left(x\right)\right)\\
q^{\left(n\right)}\!\left(t\right) & =\sum_{x}q^{\left(n\right)}\!\left(t|x\right)p\!\left(x\right)\\
 & =\sum_{x:f^{\left(n\right)}\!\left(x\right)=t}p\!\left(x\right)\\
q^{\left(n\right)}\!\left(y|t\right) & =\frac{1}{q^{\left(n\right)}\!\left(t\right)}\sum_{x}q^{\left(n\right)}\!\left(t\mid x\right)p\!\left(x,y\right)\\
 & =\frac{\sum_{x:f^{\left(n\right)}\!\left(x\right)=t}p\!\left(x,y\right)}{\sum_{x:f^{\left(n\right)}\!\left(x\right)=t}p\!\left(x\right)}.
\end{align}

This process is summarized in Algorithm~\ref{alg:DIB}.

Note that the DIB algorithm also corresponds to ``clamping'' IB
at every step by assigning each $x$ to its highest probability cluster
$t$. We can see this by taking the $\text{argmax}$ of the logarithm
of $q\!\left(t\mid x\right)$ in eqn~\ref{eq:IB_encoding}, noting
that the $\text{argmax}$ of a positive function is equivalent the
$\text{argmax}$ of its logarithm, discarding the $\log\!\left(Z\!\left(x,\beta\right)\right)$
term since it doesn't depend on $t$, and seeing that the result corresponds
to eqn~\ref{eq:DIB_encoder}. We emphasize, however, that this is
not the same as simply running the IB algorithm to convergence and
then clamping the resulting encoder; that would, in most cases, produce
a suboptimal, ``unconverged'' deterministic solution.

Like with the IB, the DIB solutions can be plotted as a function of
$\beta$. However, in this case, it is more natural to plot $I\!\left(T;Y\right)$
as a function of $H\!\left(T\right)$, rather than $I\!\left(X;T\right)$.
That said, in order to compare the IB and DIB, they need to be plotted
in the same plane. When plotting in the $I\!\left(X;T\right)$ plane,
the DIB curve will of course lie below the IB curve, since in this
plane, the IB curve is optimal; the opposite will be true when plotting
in the $H\!\left(T\right)$ plane. Comparisons with experimental data
can be performed in either plane.

\begin{algorithm}[H]
\begin{algorithmic}[1]
\State{Given $p\!\left(x,y\right)$, $\beta \geq 0$}
\State{Initialize $f^{\left(0\right)}\!\left(x\right)$}
\State{Set $q^{\left(0\right)}\!\left(t\right)=\sum_{x:f^{\left(0\right)}\!\left(x\right)=t}p\!\left(x\right)$}
\State{Set $q^{\left(0\right)}\!\left(y\mid t\right)=\frac{\sum_{x:f^{\left(0\right)}\!\left(x\right)=t}p\left(x,y\right)}{\sum_{x:f^{\left(0\right)}\!\left(x\right)=t}p\left(x\right)}$} 
\State{$n=0$}
\While{not converged}
\State{$n=n+1$}
\State{$d^{\left(n-1\right)}\!\left(x,t\right)\equiv D_{\text{KL}}\!\left[p\!\left(y\mid x\right)\mid q^{\left(n-1\right)}\!\left(y\mid t\right)\right]$}
\State{$\ell_{\beta}^{\left(n-1\right)}\!\left(x,t\right)\equiv\log q\!\left(t\right)-\beta d^{\left(n-1\right)}\!\left(x,t\right)$}
\State{$f^{\left(n\right)}\!\left(x\right)=\underset{t}{\text{argmax}}\,\,\ell_{\beta}^{\left(n-1\right)}\!\left(x,t\right)$} \State{$q^{\left(n\right)}\!\left(t\right)=\sum_{x:f^{\left(n\right)}\!\left(x\right)=t}p\!\left(x\right)$} 
\State{$q^{\left(n\right)}\!\left(y\mid t\right)=\frac{\sum_{x:f^{\left(n\right)}\!\left(x\right)=t}p\left(x,y\right)}{\sum_{x:f^{\left(n\right)}\!\left(x\right)=t}p\left(x\right)}$} \EndWhile 
\end{algorithmic}

\caption{\textbf{- The deterministic information bottleneck (DIB) method}.\label{alg:DIB}}
\end{algorithm}

\section{Comparison of IB and DIB}

To get an idea of how the IB and DIB solutions differ in practice,
we generated a series of random joint distributions $p\!\left(x,y\right)$,
solved for the IB and DIB solutions for each, and compared them in
both the IB and DIB plane. To generate the $p\!\left(x,y\right)$,
we first sampled $p\!\left(x\right)$ from a symmetric Dirichlet distribution
with concentration parameter $\alpha_{x}$ (so $p\!\left(x\right)\sim\text{Dir}\!\left[\alpha_{x}\right]$),
and then sampled each row of $p\!\left(y\mid x\right)$ from another
symmetric Dirichlet distribution with concentration parameter $\alpha_{y}^{\left(i\right)}$
(so $p\!\left(y\mid x_{i}\right)\sim\text{Dir}\!\left[\alpha_{y}^{\left(i\right)}\right]$).
In the experiments shown here, we set $\alpha_{x}$ to 1000, so that
each $x_{i}$ was approximately equally likely, and we set $\alpha_{y}^{\left(i\right)}$
to be equally spaced logarithmically between $10^{-1.3}$ and $10^{1.3}$,
in order to provide a range of informativeness in the conditionals.
We set the cardinalities of $X$ and $Y$ to $\left|X\right|=256$
and $\left|Y\right|=32$, with $\left|X\right|>\left|Y\right|$ for
two reasons. First, this encourages overlap between the conditionals
$p\!\left(y|x\right)$, which leads to a more interesting clustering
problem. Second, in typical applications, this will be the case, such
as in document clustering where there are often many more documents
that vocabulary words. Since the number of clusters in use for both
IB and DIB can only decrease from iteration to iteration (see footnote~\ref{fn:decreasingclusters-2}),
we always initialized $\left|T\right|=\left|X\right|$.\footnote{An even more efficient setting might be to set the cardinality of
$T$ based on the entropy of $X$, say $\left|T\right|=\text{ceiling}\!\left(\exp\!\left(H\!\left(X\right)\right)\right)$,
but we didn't experiment with this.} For the DIB, we initialized the cluster assignments to be as even
across the cluster as possible, i.e. each data points belonged to
its own cluster. For IB, we initialized using a soft version of the
same procedure, with 75\% of each conditional's probability mass assigned
to a unique cluster, and the remaining 25\% a normalized uniform random
vector over the remaining $\left|T\right|-1$ clusters.

\begin{figure}[H]
\begin{centering}
\includegraphics[scale=0.45]{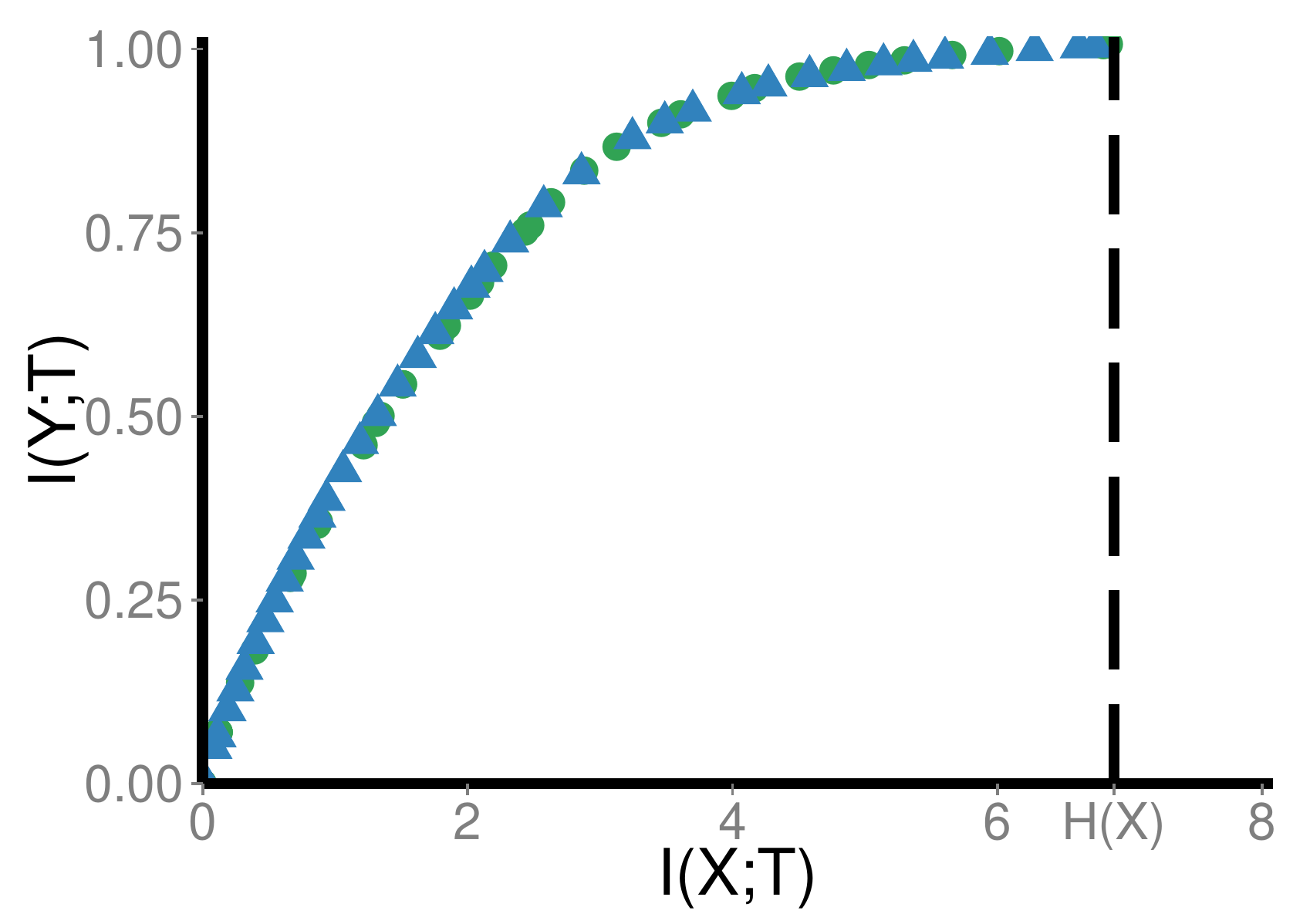}\includegraphics[scale=0.45]{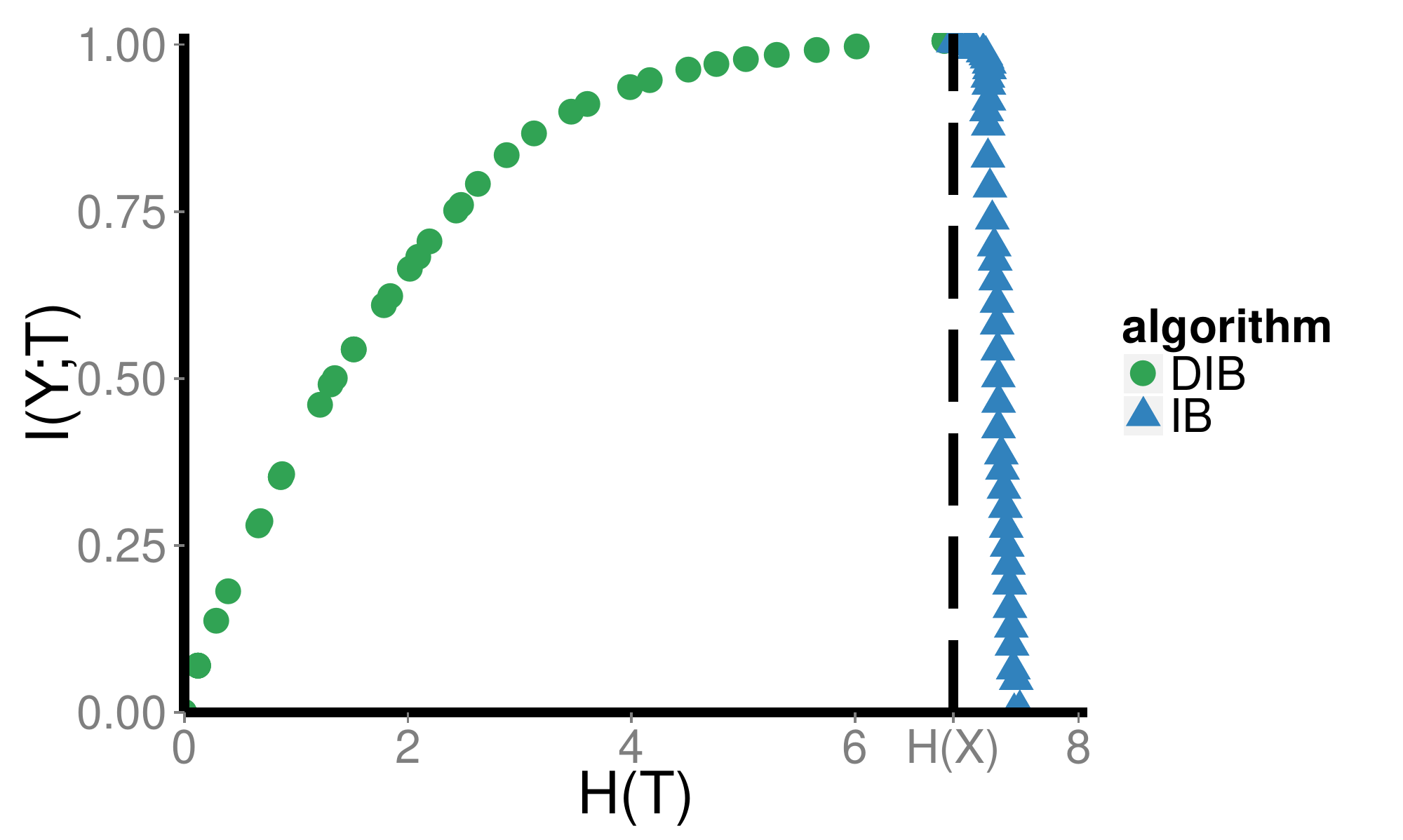} 
\par\end{centering}
\caption{\textbf{\label{fig:IB-vs-DIB}Example IB and DIB solutions.} \emph{Left}:
IB plane. \emph{Right}: DIB plane. Upper limit of the $y$-axes is
$I\!\left(X,Y\right)$, since this is the maximal possible value of
$I\!\left(T;Y\right)$. Upper limit of the $x$-axes is $\log\!\left(\left|T\right|\right)$,
since this is the maximal possible value of $H\!\left(T\right)$ and
$I\!\left(X,T\right)$ (the latter being true since $I\!\left(X,T\right)$
is bounded above by both $H\!\left(T\right)$ and $H\!\left(X\right)$,
and $\left|T\right|<\left|X\right|$). The dashed vertical lines mark
$H\!\left(X\right)$, which is both an upper bound for $I\!\left(X,T\right)$
and a natural comparison for $H\!\left(T\right)$ (since to place
each data point in its own cluster, we need $H\!\left(T\right)=H\!\left(X\right)$).}
\end{figure}

An illustrative pair of solutions is shown in figure~\ref{fig:IB-vs-DIB}.
The key feature to note is that, while performance of the IB and DIB
solutions are very similar in the IB plane, their behavior differs
drastically in the DIB plane.

Perhaps most unintuitive is the behavior of the IB solution in the
DIB plane, where from an entropy perspective, the IB actually ``decompresses''
the data (i.e. $H\!\left(T\right)>H\!\left(X\right)$). To understand
this behavior, recall that the IB's compression term is the mutual
information $I\!\left(X,T\right)=H\!\left(T\right)-H\!\left(T\mid X\right)$.
This term is minimized by any $q\!\left(t\mid x\right)$ that maps
$t$s independently of $x$s. Consider two extremes of such mappings.
One is to map all values of $X$ to a single value of $T$; this leads
to $H\!\left(T\right)=H\!\left(T\mid X\right)=I\!\left(X,T\right)=0$.
The other is to map each value of $X$ uniformly across all values
of $T$; this leads to $H\!\left(T\right)=H\!\left(T\mid X\right)=\log\left|T\right|$
and $I\!\left(X,T\right)=0$. In our initial studies, the IB consistently
took the latter approach.\footnote{Intuitively, this approach is ``more random'' and is therefore easier
to stumble upon during optimization.} Since the DIB cost function favors the former approach (and indeed
the DIB solution follows this approach), the IB consistently performs
poorly by the DIB's standards. This difference is especially apparent
at small $\beta$, where the compression term matters most, and as
$\beta$ increases, the DIB and IB solutions converge in the DIB plane.

\begin{figure}[H]
\begin{centering}
\includegraphics[scale=0.45]{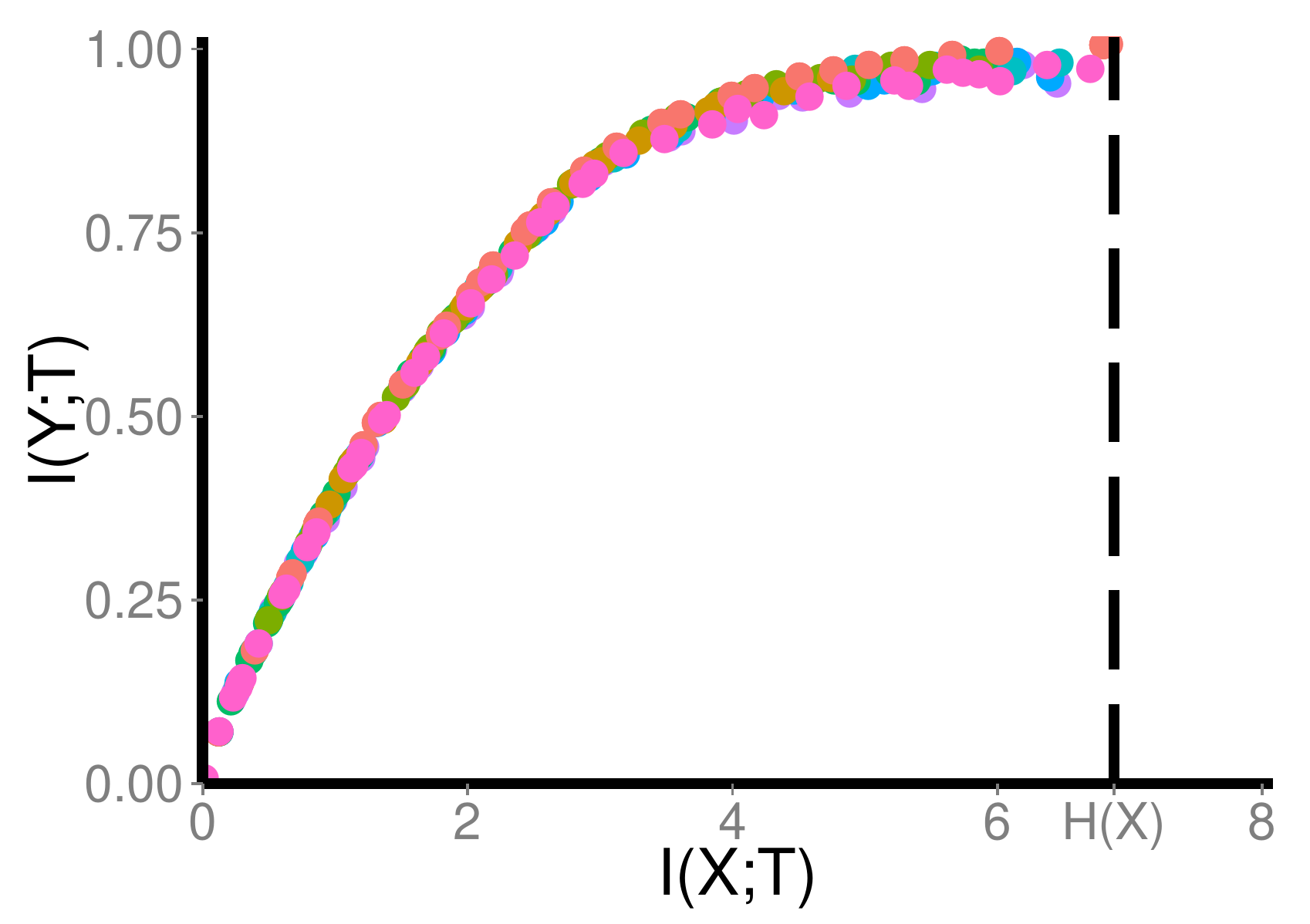}\includegraphics[scale=0.45]{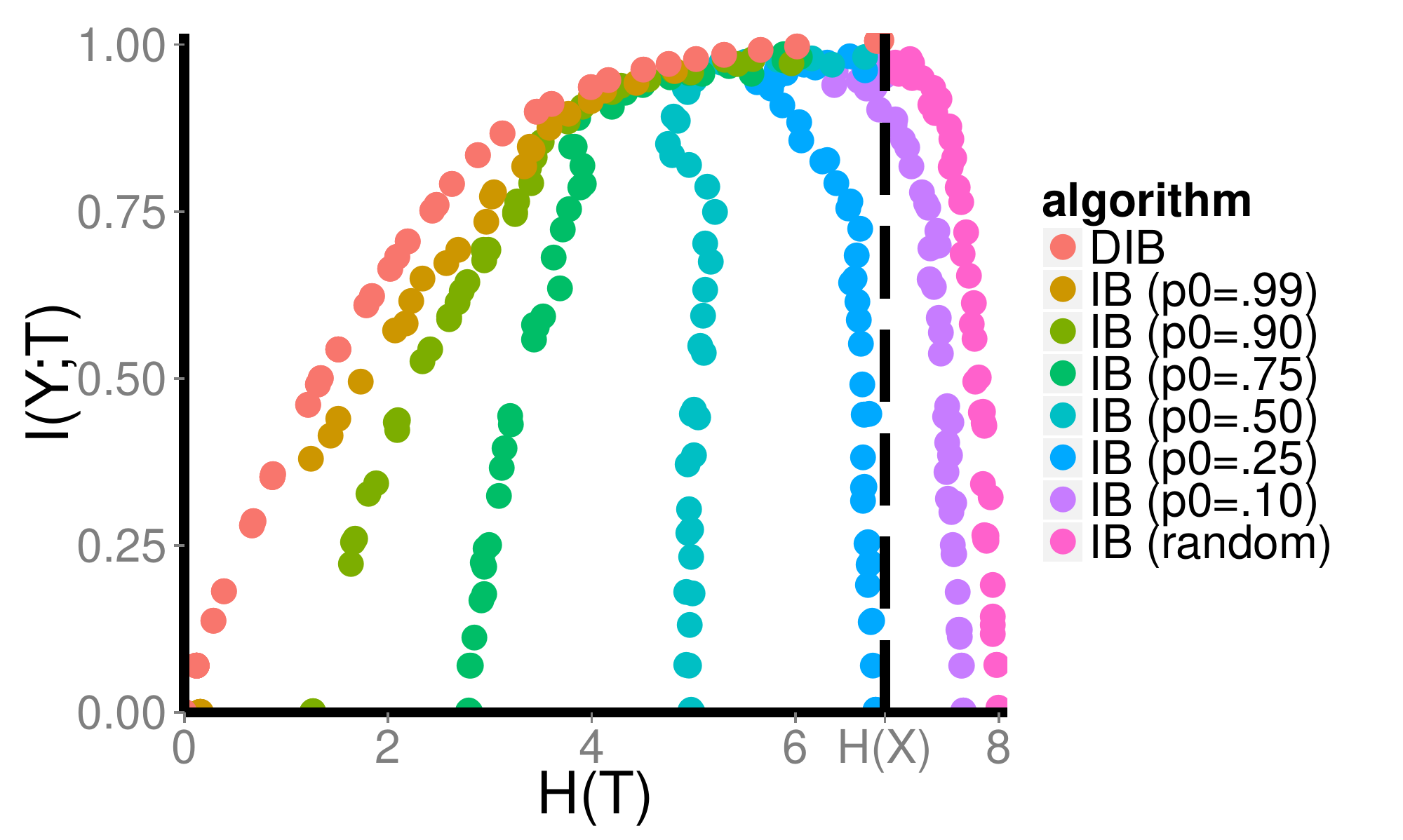} 
\par\end{centering}
\caption{\textbf{\label{fig:IB-vs-DIB-diffinits}Example IB and DIB solutions
across different IB initializations.}\emph{ }Details identical to
Figure~\ref{fig:IB-vs-DIB}, except colors represent different initializations
for the IB, as described in the text.}
\end{figure}

To encourage the IB to perform closer to DIB optimality at small $\beta$,
we next altered our initialization scheme of $q\!\left(t\mid x\right)$
to one biased towards single-cluster solutions; instead of each $x_{i}$
having most of its probability mass on a unique cluster $t_{i}$,
we placed most of the probability mass for each $x_{i}$ on the \emph{same}
cluster $t^{*}$. The intended effect was to start the IB closer to
solutions in which all data points were mapped to a single cluster.
Results are shown in figure~\ref{fig:IB-vs-DIB-diffinits}. Here,
$p_{0}$ is the amount of probability mass placed on the cluster $t^{*}$,
that is $q\!\left(t^{*}\mid x\right)=p_{0},\thinspace\thinspace\forall x$;
the probability mass for the remaining $\left|T\right|-1$ clusters
was again initialized to a normalized uniform random vector. ``random''
refers to an initialization which skips placing the $p_{0}$ mass
and just initializes each $q\!\left(t\mid x_{i}\right)$ to a normalized
uniform random vector.

We note several features. First, although we can see a gradual shift
of the IB towards DIB-like behavior in the DIB plane as $p_{0}\rightarrow1$,
the IB solutions never quite reach the performance of DIB. Moreover,
as $p_{0}\rightarrow1$, the single-cluster initializations exhibit
a phase transition in which, regardless of $\beta$, they ``skip''
over a sizable fraction of lower-$I\!\left(Y;T\right)$ solutions
that are picked up by DIB. Third, even for higher-$I\!\left(Y;T\right)$
solutions, the single-cluster initializations seem to perform suboptimally,
not quite extracting all of the information $I\!\left(X;Y\right)$,
as DIB and the multi-cluster initialization from the previous section
do; this can be seen in both the IB and DIB plane.

To summarize, the IB and DIB perform similarly by the IB standards,
but the DIB tends to outperform the IB dramatically by the DIB's standards.
Careful initialization of the IB can make up some of the difference,
but not all.

\begin{figure}[H]
\begin{centering}
\includegraphics[scale=0.45]{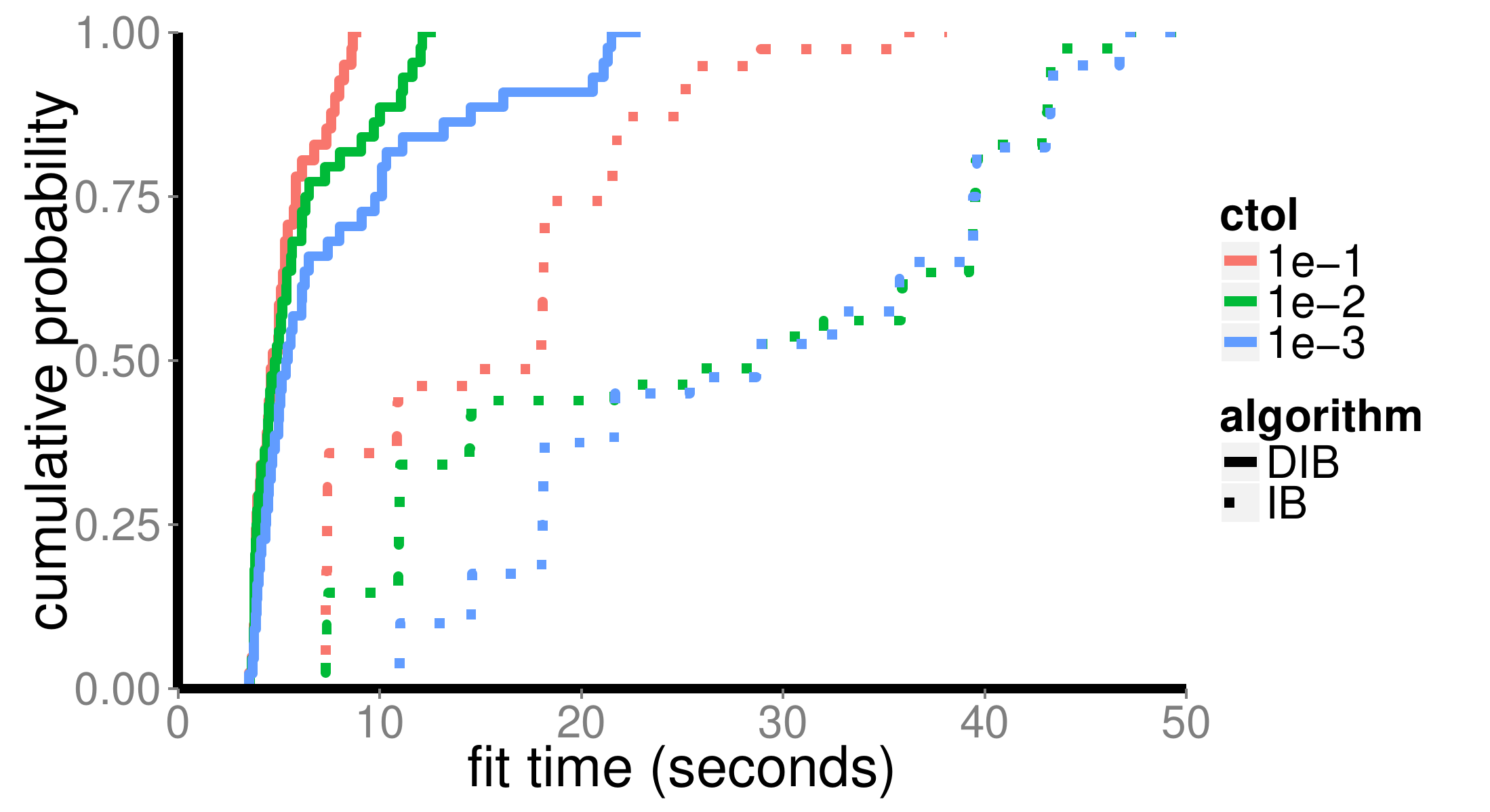}
\par\end{centering}
\caption{\textbf{\label{fig:IB-vs-DIB-fittimes}Fit times for IB and DIB.}
Cumulative distribution function of fit times across $\beta$, for
a variety of settings of the convergence tolerance. Note that absolute
numbers here depend on hardware, so we emphasize only relative comparisons
of IB vs DIB. Note also that across the range of $\text{ctol}$ values
we tested here, the (D)IB curves vary by less than the width of the
data points, and so we omit them.}
\end{figure}

It is also worth noting that, across all the datasets we tested, the
DIB also tended to converge faster, as illustrated in figure~\ref{fig:IB-vs-DIB-fittimes}.
The DIB speedup over IB varied depended on the convergence conditions.
In our experiments, we defined convergence as when the relative step-to-step
change in the cost functional $L$ was smaller than some threshold
$\text{ctol}$, that is when $\frac{L_{n-1}-L_{n}}{L_{n-1}}<\text{ctol}$
at step $n$. In the results above, we used $\text{ctol}=10^{-3}$.
In figure~\ref{fig:IB-vs-DIB-fittimes}, we vary $\text{ctol}$,
with the IB initialization scheme fixed to the original ``multi-cluster''
version, to show the effect on the relative speedup of DIB over IB.
While DIB remained $\approx$2-5x faster than IB in all cases tested,
that speedup tended to be more pronounced with lower $\text{ctol}$.
Since the ideal convergence conditions would probably vary by dataset
size and complexity, it is difficult to make any general conclusions,
though our experiments do at least suggest that DIB offers a computational
advantage over IB.

\section{Related work\label{sec:Related-work}}

The DIB is not the first hard clustering version of IB.\footnote{In fact, even the IB itself produces a hard clustering in the large
$\beta$ limit. However, it trivially assigns all data points to their
own clusters.} Indeed, the agglomerative information bottleneck (AIB) \cite{AIB}
also produces hard clustering and was introduced soon after the IB.
Thus, it is important to distinguish between the two approaches. AIB
is a bottom-up, greedy method which starts with all data points belonging
to their own clusters and iteratively merges clusters in a way which
maximizes the gain in relevant information. It was explicitly designed
to produce a hard clustering. DIB is a top-down method derived from
a cost function that was not designed to produce a hard clustering.
Our starting point was to alter the IB cost function to match the
source coding notion of compression. The emergence of hard clustering
in DIB is itself a result. Thus, while AIB does provide a hard clustering
version of IB, DIB contributes the following in addition: 1) Our study
emphasizes why a stochastic encoder is optimal for IB, namely due
to the noise entropy term. 2) Our study provides a principled, top-down
derivation of a hard clustering version of IB, based upon an intuitive
change to the cost function. 3) Our non-trivial derivation also provides
a cost function and solution which interpolates between DIB and IB,
by adding back the noise entropy continuously, i.e. with $0<\alpha<1$.
This interpolation may be viewed as adding a regularization term to
DIB, one that may perhaps be useful in dealing with finitely sampled
data. Another interpretation of the cost function with intermediate
$\alpha$ is as a penalty on \emph{both }the mutual information between
$X$ and $T$ \emph{and} the entropy of the compression, $H(T)$.

The original IB also provides a deterministic encoding upon taking
the limit $\beta\rightarrow\infty$ that corresponds to the causal-state
partition of histories \cite{Crutch}. However, this is the limit
of no compression, whereas our approach allows for an entire family
of deterministic encoders with varying degrees of compression.

\section{Discussion}

Here we have introduced the deterministic information bottleneck (DIB)
as an alternative to the information bottleneck (IB) for compression
and clustering. We have argued that the DIB cost function better embodies
the goal of lossy compression of relevant information, and shown that
it leads to a non-trivial deterministic version of the IB. We have
compared the DIB and IB solutions on synthetic data and found that,
in our experiments, the DIB performs nearly identically to the IB
in terms of the IB cost function, but far superior in terms of its
own cost function. We also noted that the DIB achieved this performance
at a computational efficiency 2-5x better than the IB.

Of course, in addition to the studies with synthetic data here, it
is important to compare the DIB and IB on real world datasets as well
to see whether the DIB's apparent advantages hold, for example with
datasets that have more explicit hierarchical structure for both algorthms
to exploit, such as in topic modelling \cite{HierarchicalTopicModel,ML-IB}.

One particular application of interest is maximally informative clustering,
where it would be interesting to know how IB and DIB relate to classic
clustering algorithms such as $k$-means \cite{geometric}. Previous
work has, for example, offered a principled way of choosing the number
of clusters based on the finiteness of the data \cite{Clusters},
and similarly interesting results may exist for the DIB. More generally,
there are learning theory results showing generalization bounds on
IB for which an analog on DIB would be interesting as well \cite{Generalization}.

\begin{sloppypar}Another potential area of application is modeling
the extraction of predictive information in the brain (which is one
particular example in a long line of work on the exploitation of environmental
statistics by the brain \cite{Barlow1,Barlow2,Barlow3,Atick,OlshausenField1,OlshausenField2,SimoncelliOlshausen,OlshausenField3}).
There, $X$ would be the stimulus at time $t$, $Y$ the stimulus
a short time in the future $t+\tau$, and $T$ the activity of a population
of sensory neurons. One could even consider neurons deeper in the
brain by allowing $X$ and $Y$ to correspond not to an external stimulus,
but to the activity of upstream neurons. An analysis of this nature
using retinal data was recently performed with the IB \cite{Palmer}.
It would be interesting to see if the same data corresponds better
to the behavior of the DIB, particularly in the DIB plane where the
IB and DIB differ dramatically.\end{sloppypar}

We close by noting that DIB is an imperfect name for the algorithm
introduced here for a couple of reasons. First, there do exist other
deterministic limits and approximations to the IB (see, for example,
the discussion of the AIB in section~\ref{sec:Related-work}), and
so we hesitate to use the phrase ``the'' deterministic IB. Second,
our motivation here was not to create a deterministic version of IB,
but instead to alter the cost function in a way that better encapsulates
the goals of certain problems in data analysis. Thus, the deterministic
nature of the solution was a result, not a goal. For this reason,
``entropic bottleneck'' might also be an appropriate name.

\section{Acknowledgements}

For insightful discussions, we would like to thank Richard Turner,
Máté Lengyel, Bill Bialek, Stephanie Palmer, and Gordon Berman. We
would also like to acknowledge financial support from NIH K25 GM098875
(Schwab), the Hertz Foundation (Strouse), and the Department of Energy
Computational Sciences Graduate Fellowship (Strouse).

\section{Appendix: derivation of generalized IB solution\label{sec:Derivation}}

Given $p\!\left(x,y\right)$ and subject to the Markov constraint
$T\leftrightarrow X\leftrightarrow Y$, the generalized IB problem
is: 
\begin{align}
\min_{q\left(t|x\right)}L\!\left[q\!\left(t|x\right)\right] & =H\!\left(T\right)-\alpha H\!\left(T|X\right)-\beta I\!\left(Y;T\right)-\sum_{x,t}\lambda\!\left(x\right)q\!\left(t|x\right),
\end{align}

where we have now included the Lagrange multiplier term (which enforces
normalization of $q\!\left(t|x\right)$) explicitly. The Markov constraint
implies the following factorizations: 
\begin{align}
q\!\left(t|y\right) & =\sum_{x}q\!\left(t|x\right)p\!\left(x|y\right)\\
q\!\left(t\right) & =\sum_{x}q\!\left(t|x\right)p\!\left(x\right),
\end{align}

which give us the following useful derivatives: 
\begin{align}
\frac{\delta q\!\left(t|y\right)}{\delta q\!\left(t|x\right)} & =p\!\left(x|y\right)\\
\frac{\delta q\!\left(t\right)}{\delta q\!\left(t|x\right)} & =p\!\left(x\right).
\end{align}

Now taking the derivative of the cost function with respect to the
encoding distribution, we get: 
\begin{align}
\frac{\delta L}{\delta q\!\left(t|x\right)} & =-\frac{\delta}{\delta q\!\left(t|x\right)}\sum_{t}q\!\left(t\right)\log q\!\left(t\right)-\frac{\delta}{\delta q\!\left(t|x\right)}\sum_{x,t}\lambda\!\left(x\right)q\!\left(t|x\right)\\
 & +\alpha\frac{\delta}{\delta q\!\left(t|x\right)}\sum_{x,t}q\!\left(t|x\right)p\!\left(x\right)\log q\!\left(t|x\right)\\
 & -\beta\frac{\delta}{\delta q\!\left(t|x\right)}\sum_{y,t}q\!\left(t|y\right)p\!\left(y\right)\log\!\left[\frac{q\!\left(t|y\right)}{q\!\left(t\right)}\right]\\
 & =-\log q\!\left(t\right)\frac{\delta q\!\left(t\right)}{\delta q\!\left(t|x\right)}-q\!\left(t\right)\frac{\delta\log q\!\left(t\right)}{\delta q\!\left(t|x\right)}-\lambda\!\left(x\right)\frac{\delta q\!\left(t|x\right)}{\delta q\!\left(t|x\right)}\\
 & +\alpha\left[p\!\left(x\right)\log q\!\left(t|x\right)\frac{\delta q\!\left(t|x\right)}{\delta q\!\left(t|x\right)}+q\!\left(t|x\right)p\!\left(x\right)\frac{\delta\log q\!\left(t|x\right)}{\delta q\!\left(t|x\right)}\right]\\
 & -\beta\sum_{y}\left[p\!\left(y\right)\log\!\left[\frac{q\!\left(t|y\right)}{q\!\left(t\right)}\right]\frac{\delta q\!\left(t|y\right)}{\delta q\!\left(t|x\right)}\right]\\
 & +\beta\sum_{y}\left[q\!\left(t|y\right)p\!\left(y\right)\frac{\delta\log q\!\left(t|y\right)}{\delta q\!\left(t|x\right)}+q\!\left(t|y\right)p\!\left(y\right)\frac{\delta\log q\!\left(t\right)}{\delta q\!\left(t|x\right)}\right]\\
 & =-p\!\left(x\right)\log q\!\left(t\right)-p\!\left(x\right)-\lambda\!\left(x\right)+\alpha\left[p\!\left(x\right)\log q\!\left(t|x\right)+p\!\left(x\right)\right]\\
 & -\beta\sum_{y}\left[p\!\left(y\right)\log\!\left[\frac{q\!\left(t|y\right)}{q\!\left(t\right)}\right]p\!\left(x|y\right)+p\!\left(y\right)p\!\left(x|y\right)-q\!\left(t|y\right)p\!\left(y\right)\frac{p\!\left(x\right)}{q\!\left(t\right)}\right]\\
 & =-p\!\left(x\right)\log q\!\left(t\right)-p\!\left(x\right)-\lambda\!\left(x\right)+\alpha\left[p\!\left(x\right)\log q\!\left(t|x\right)+p\!\left(x\right)\right]\\
 & -\beta p\!\left(x\right)\left[\sum_{y}p\!\left(y|x\right)\log\!\left[\frac{q\!\left(t|y\right)}{q\!\left(t\right)}\right]+\sum_{y}p\!\left(y|x\right)-\sum_{y}q\!\left(y|t\right)\right]\\
 & =p\!\left(x\right)\left[-1-\log q\!\left(t\right)-\frac{\lambda\!\left(x\right)}{p\!\left(x\right)}+\alpha\log q\!\left(t|x\right)+\alpha-\beta\left[\sum_{y}p\!\left(y|x\right)\log\!\left[\frac{q\!\left(t|y\right)}{q\!\left(t\right)}\right]\right]\right].
\end{align}

Setting this to zero implies that: 
\begin{align}
\alpha\log q\!\left(t|x\right) & =1-\alpha+\log q\!\left(t\right)+\frac{\lambda\!\left(x\right)}{p\!\left(x\right)}+\beta\left[\sum_{y}p\!\left(y|x\right)\log\!\left[\frac{q\!\left(t|y\right)}{q\!\left(t\right)}\right]\right].
\end{align}

We want to rewrite the $\beta$ term as a KL divergence. First, we
will need that $\log\!\left[\frac{q\left(t|y\right)}{q\left(t\right)}\right]=\log\!\left[\frac{q\left(t,y\right)}{q\left(t\right)p\left(y\right)}\right]=\log\!\left[\frac{q\left(y|t\right)}{p\left(y\right)}\right]$.
Second, we will add and subtract $\beta\sum_{y}p\!\left(y|x\right)\log\!\left[\frac{p\left(y|x\right)}{p\left(y\right)}\right]$.
This gives us: 
\begin{align}
\alpha\log q\!\left(t|x\right) & =1-\alpha+\log q\!\left(t\right)+\frac{\lambda\!\left(x\right)}{p\!\left(x\right)}+\beta\sum_{y}p\!\left(y|x\right)\log\!\left[\frac{p\!\left(y|x\right)}{p\!\left(y\right)}\right]\\
 & -\beta\left[\sum_{y}p\!\left(y|x\right)\log\!\left[\frac{p\!\left(y|x\right)}{q\!\left(y|t\right)}\right]\right].
\end{align}

The second $\beta$ term is now just $D_{\text{KL}}\!\left[p\!\left(y|x\right)\mid q\!\left(y|t\right)\right]$.
Dividing both sides by $\alpha$, this leaves us with the equation:
\begin{align}
\log q\!\left(t|x\right) & =z\!\left(x,\alpha,\beta\right)+\frac{1}{\alpha}\log q\!\left(t\right)-\frac{\beta}{\alpha}D_{\text{KL}}\!\left[p\!\left(y\mid x\right)|q\!\left(y|t\right)\right],
\end{align}

where we have absorbed all of the terms that don't depend on $t$
into a single factor:
\begin{align}
z\!\left(x,\alpha,\beta\right) & \equiv\frac{1}{\alpha}-1+\frac{\lambda\!\left(x\right)}{\alpha p\!\left(x\right)}+\frac{\beta}{\alpha}\sum_{y}p\!\left(y\mid x\right)\log\!\left[\frac{p\!\left(y\mid x\right)}{p\!\left(y\right)}\right].
\end{align}

Solving for $q\!\left(t|x\right)$, we get: 
\begin{align}
q\!\left(t|x\right) & =\exp\!\left[z\right]\exp\!\left[\frac{1}{\alpha}\left(\log q\!\left(t\right)-\beta D_{\text{KL}}\!\left[p\!\left(y|x\right)\mid q\!\left(y|t\right)\right]\right)\right]\\
 & =\frac{1}{Z}\exp\!\left[\frac{1}{\alpha}\left(\log q\!\left(t\right)-\beta D_{\text{KL}}\!\left[p\!\left(y|x\right)\mid q\!\left(y|t\right)\right]\right)\right],
\end{align}

where:

\begin{align}
Z\!\left(x,\alpha,\beta\right) & \equiv\exp\!\left[-z\right]
\end{align}

is just a normalization factor. Now that we're done with the general
derivation, let's add a subscript to the solution to distinguish it
from the special cases of the IB and DIB. 
\begin{align}
q_{\alpha}\!\left(t|x\right) & =\frac{1}{Z\!\left(x,\alpha,\beta\right)}\exp\!\left[\frac{1}{\alpha}\left(\log q\!\left(t\right)-\beta D_{\text{KL}}\!\left[p\!\left(y|x\right)\mid q\!\left(y|t\right)\right]\right)\right].
\end{align}

The IB solution is then: 
\begin{align}
q_{\text{IB}}\!\left(t|x\right)=q_{\alpha=1}\!\left(t|x\right) & =\frac{q\!\left(t\right)}{Z}\exp\!\left[-\beta D_{\text{KL}}\!\left[p\!\left(y|x\right)\mid q\!\left(y|t\right)\right]\right],
\end{align}

while the DIB solution is: 
\begin{align}
q_{\text{DIB}}\!\left(t|x\right)=\lim_{\alpha\rightarrow0}q_{\alpha}\!\left(t|x\right) & =\delta\!\left(t-t^{*}\!\left(x\right)\right),
\end{align}

with: 
\begin{align}
t^{*}\!\left(x\right) & =\text{\ensuremath{\underset{t}{\operatorname{argmax}}}}\left(\log q\!\left(t\right)-\beta D_{\text{KL}}\!\left[p\!\left(y|x\right)\mid q\!\left(y|t\right)\right]\right).
\end{align}

\end{document}